\documentclass[aps,twocolumn,showpacs,superscriptaddress]{revtex4-1}
\usepackage{natbib}
\usepackage{amsfonts,amsmath,amssymb,graphics,graphicx,dcolumn,bm,multirow,enumerate,verbatim}
\usepackage{hyperref}

\newcommand{\cu}
{\affiliation{Department of Physics, University of Calcutta, 
92 Acharya Prafulla Chandra Road, Kolkata 700009, India.}}
\newcommand{\aalto}
{\affiliation{Department of Biomedical Engineering and Computational Science, 
Aalto University School of Science, P.O.  Box 12200, FI-00076, Finland}}

\begin{document}

\title
{On the evolution and utility of annual citation indices}

\author{Abdul Khaleque}
\email[Email: ]{7187ak@gmail.com}
\cu
\author{Arnab Chatterjee}%
\email[Email: ]{arnab.chatterjee@aalto.fi}
\aalto
\author{Parongama Sen}%
\email[Email: ]{psphy@caluniv.ac.in}
\cu

\begin{abstract}

We study the statistics of citations made to 
the top ranked indexed journals 
for Science and Social Science databases in the Journal Citation Reports using 
different measures. 
Total  annual citation and impact factor,  
 as well as a   third measure  called the annual citation rate 
 are used to make the detailed analysis. 
We observe that the distribution of the annual citation rate has an universal feature - it shows
a maximum at the rate scaled by half the average, irrespective of how the journals are ranked,
and even across Science and Social Science journals, and fits well to log-Gumbel distribution.
Correlations between different  quantities 
are studied and a comparative  analysis of the three measures is presented.
 The newly introduced annual citation rate factor helps in understanding the effect of scaling the number of citation by the 
total number of publications. 
The effect of the impact factor on  authors contributing to the journals as well as on editorial 
policies  is also discussed.
\end{abstract}
\maketitle

\section{Introduction}
The popularity of an academic journal may be related to   
its readership, while  the quality is usually
evaluated  by several factors related to the citations it receives.
Among them, the total citations in a year, the  impact factor~\cite{Garfield:1964,Garfield:2006}, 
the eigenfactors~\cite{Bergstrom:2008} are popular measures.
The \textit{impact factor} (IF)~\cite{Garfield:1964,Garfield:2006} of  
an academic journal is a measure which reflects 
the average number of citations to recent articles published in the same journal. 
It is frequently used as a proxy for the relative importance of a journal within its field, 
with journals with higher IFs deemed to be more important compared to those with lower ones. 
The \textit{eigenfactor} measure in addition takes into account the quality of the 
journals in which the citing articles appear, arguing that a journal is considered 
to be more influential if it is cited often by other influential journals. 
It was shown~\cite{Fersht:2009} however that the eigenfactor measurement 
is more or less correlated with the annual citation measure.

There have been plenty of empirical studies on citation data~\cite{Sen:2013}, specifically on 
citation distributions~\cite{Shockley:1957,Laherrere:1998,Redner:1998,Radicchi:2008} of articles,  
probability of citation as a function of time~\cite{Rousseau:1994,Egghe:2000,Burrell:2001,Burrell:2002}, citations of 
individuals authors~\cite{Petersen:2011} and their dynamics~\cite{Eom:2011}.
It has been recently shown that the $h$-index~\cite{hirsch2005index} is weakly correlated
with number of publications of a scientist, but is strongly correlated with the number of citations that one has received,
suggesting that the number of citations can be effectively used as a proxy of the $h$-index~\cite{radicchi2013analysis}.

Apart from 
studying the properties/statistics of the standard measures of annual citation
and impact factor, we also introduce and analyse a new measure
called the citation rate. 
Even when one considers a truncated dataset (as is the case here) 
this measure exhibits more characteristic features compared to the other two with respect to certain properties. 
The motivation for the present work is to study mainly
the statistical properties of annual citation, its rate and impact 
factors -- several
 distributions and  correlations are investigated for all these quantities.

\section{Quantities of interest}
Impact factors are calculated yearly for journals that are indexed in the Journal Citation Reports~\cite{JCR}.
The precise definition of IF is the following: if papers published in a journal
in years $T-2$ and $T-1$ are cited ${\mathcal {N}}(T-2) + {\mathcal {N}}(T-1)$ times by indexed journals in the year $T$, 
and $N(T-2) + N(T-1)$ be the number of citable articles published in those years, then 
the impact factor in year $T$ is given by
\begin{equation}
\label{eq:if}
I(T) = \frac{{\mathcal {N}}(T-2) + {\mathcal {N}}(T-1)}{N(T-2) + N(T-1)}.
\end{equation}
One can also measure $n(T)$, the  number of annual citations (AC)  to a journal in a given year.
This is given by 
\begin{equation}
\label{eq:cita}
n(T) = \sum_{t \leq T}{\sum_i {\mathcal{A}}_i (t,T)},
\end{equation}
where ${\mathcal {A}}_i(t,T)$ is the citations received in the year $T$ 
by the $i$ th paper published 
in the year $t \leq T$.

We introduce another measure, $r(T)$, the \textit{annual citation rate} (CR)
  at a particular year $T$ 
that is   defined as the number of citations received in a year (annual citations) 
divided by the number of articles published in the \textit{same} year. Formally, 
\begin{equation}
\label{eq:cita-rate}
r(T) = n(T)/N(T).
\end{equation}
This quantity is \textit{local} in the sense it is for the same year, and
can also be interpreted as \textit{non-local} as well, because the citations are
received for all articles published in the journal in time history (for $t \le T$).

\section{Data}
\begin{figure}
  \begin{center}
  \includegraphics[width=8.5cm]{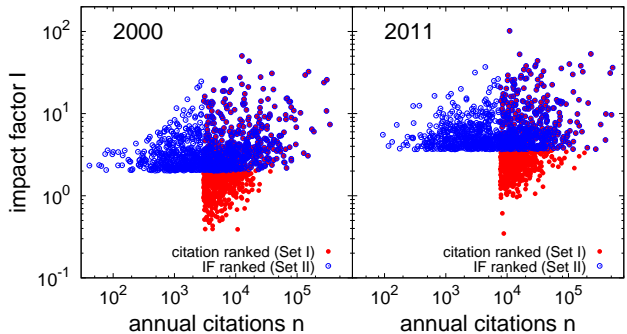}
    \end{center}
      \caption{
    Scatter plot of annual citation and impact factor. Scatter plot of citation ($n$) vs. impact factor (IF)
  for the top $1000$ journals, ranked according to citations (Set I, filled red) and impact factors (Set II, open blue).
 The plots are shown for 2000 and 2011 for comparison. The data is from SCI sets.}
\label{fig:nvsi}
 \end{figure}
We collected data for the top $1000$ journals, ranked according to 
(i) the number of citations $n(T)$ received by the journal in a year $T$ (Set I) and 
(ii) IF $I(T)$ in that year $T$ (Set II), 
for each of several years for the Science (SCI) and Social Science (SOCSCI) databases indexed
in the Journal Citation Reports (JCR)~\cite{JCR}.
We analyzed $13$ years ($2000-2012$) of data for the Science and
$6$ years ($2007-2012$) of data for the Social Science journals. 
All these data sets contained at least the information about the following quantities:
(i) the number of citations $n(T)$ received by the journal in a year, 
(ii) IF $I(T)$, (iii) the number of articles published $N(T)$ in the journal
in the same year $T$ and  a few other quantities.
Fig.~\ref{fig:nvsi} shows the journals with their citations and impact factors,
for the two datasets, ordered in different manner.
\section{Results}
\subsection{Rank plots}
\label{rank}
\begin{figure*}
 \begin{center}
 \includegraphics[width=8.6cm]{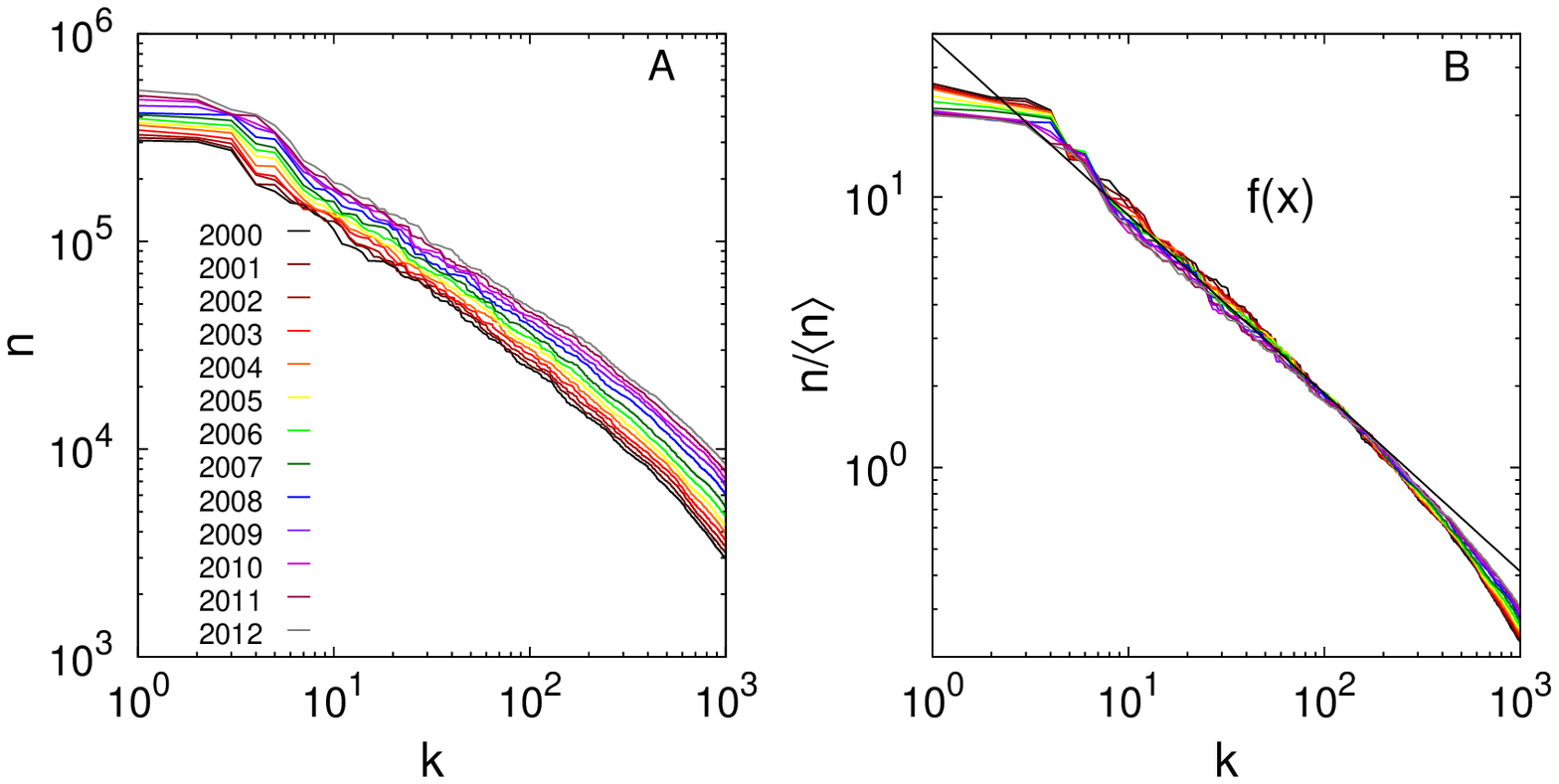}
  \includegraphics[width=8.6cm]{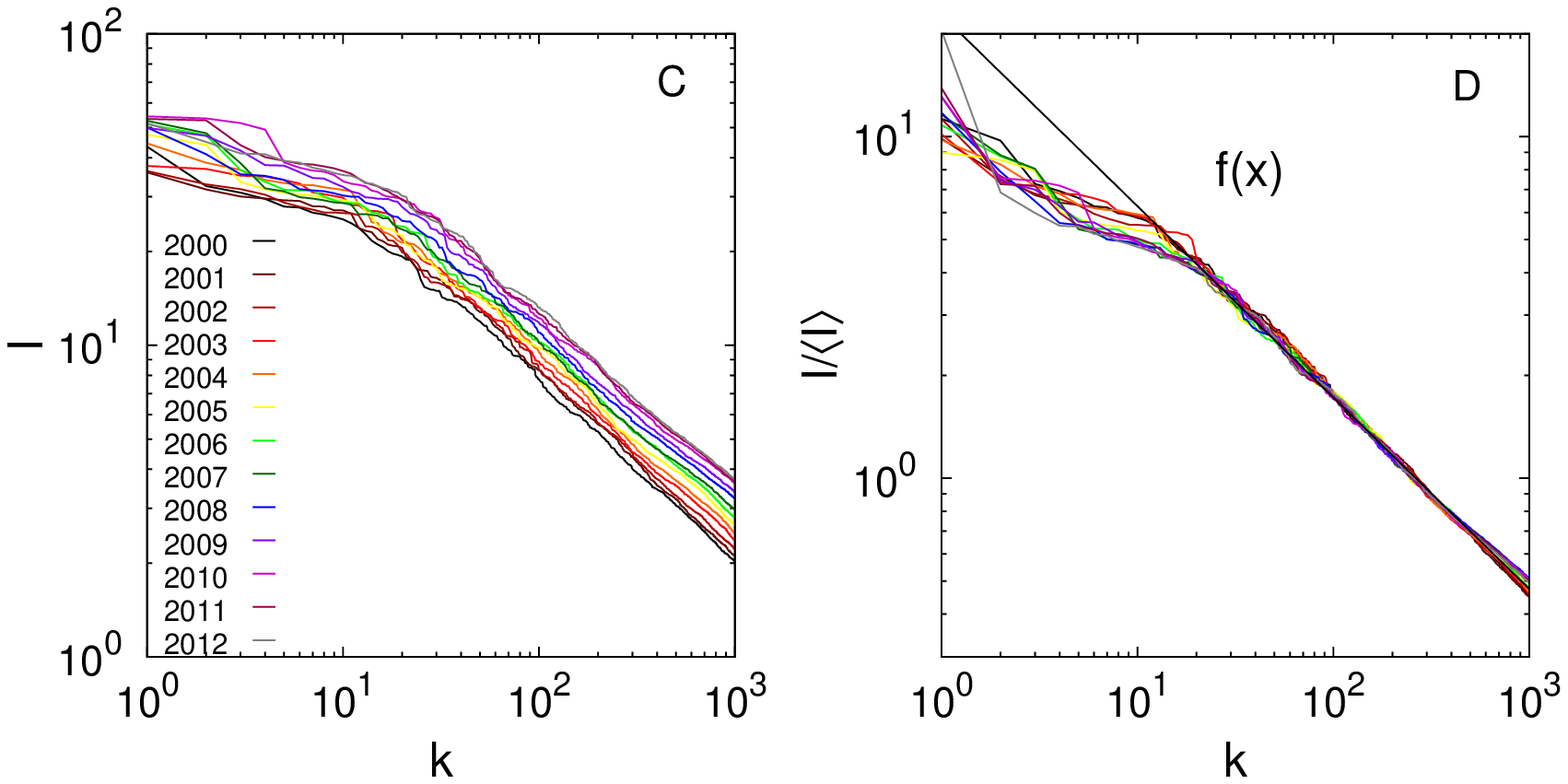}
  \includegraphics[width=8.7cm]{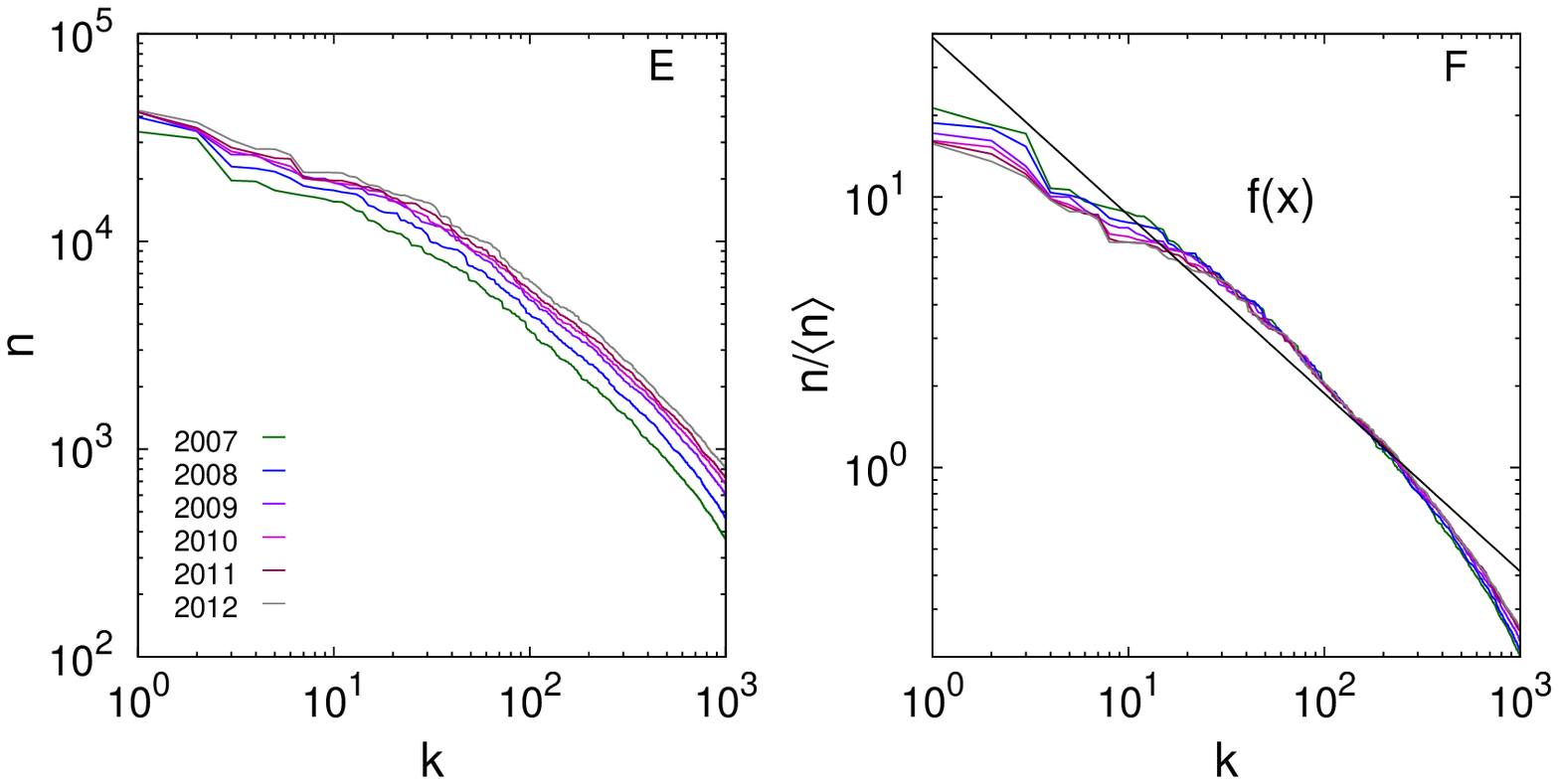}
  \includegraphics[width=8.7cm]{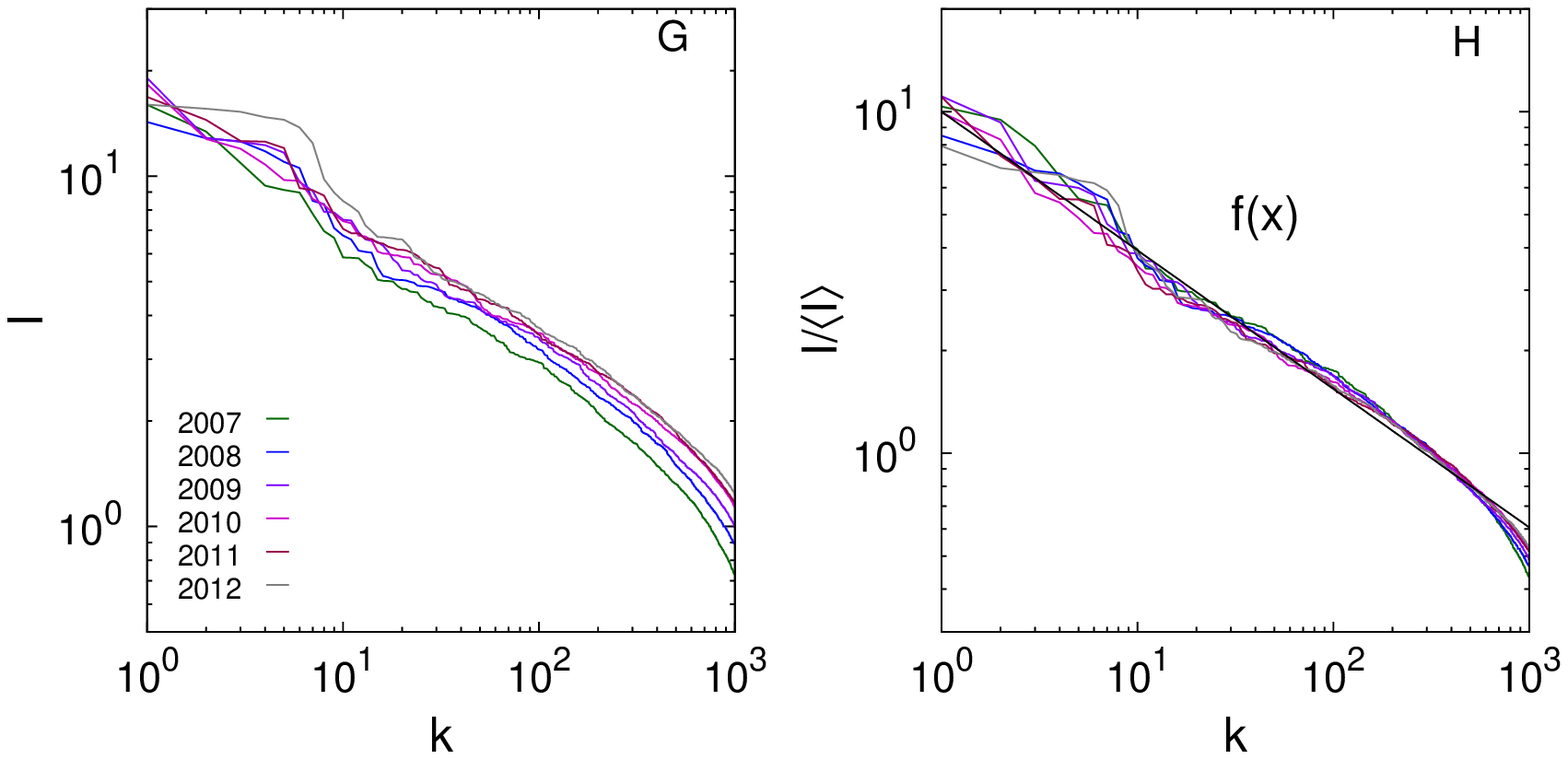}
    \end{center}
   \caption{Plots of annual citation $n$ and impact factor $I$ with rank $k$: 
   (A) Rank plot of the top $1000$ journals, ranked according to citations,
    (B) scaling collapse of the same, with a Zipf law fit: $f(x)=Ax^{-b_n}$, with 
$b_n = 0.70(2)$.
    (C) rank plot of the top $1000$ journals, ranked according to impact factors, and
 (D) scaling collapse of the same, with a Zipf law fit: $g(x)=Ax^{-b_I}$, 
with $b_I=0.54(1)$.
 The data is from SCI sets.
 SOCSCI data: Rank plots. (E) Rank plot of the top $1000$ journals, ranked according to citations,
    (F) scaling collapse of the same, with a Zipf law fit: $f(x)=Ax^{-b_n}$, 
with 
 $b_n=
0.7043 \pm 0.001$
    (G) rank plot of the top $1000$ journals, ranked according to their impact factors, and
 (H) scaling collapse of the same, with a Zipf law fit: $f(x)=Ax^{-b_I}$, 
$b_I= 0.40(1)$.}
\label{fig:rank}
\end{figure*}
To begin with, we looked at the ranked plot for both Set I and II.
Fig.~\ref{fig:rank} shows the  values of (i) AC $n$
and (ii) IF  $I$ against their rank, ranked
according to the values of the same quantities.
On the  whole, a journal at a rank $k$ is observed to increase its AC and IF over years. However, intense competition among top ranked journals
is apparent from the occasional crossing of the curves for different years for 
the highest ranks ($k < 10$). For ranks $k > 100$, the behavior is rather regular, 
in the sense that these curves never or rarely cross. 
Extensive studies on the historical behavior of the IF ranked 
distribution~\cite{Popescu:2003,Mansilla:2007} have established this behavior. These historical
studies concentrate on the behavior of the low ranked (large $k$) journals
and the precise nature of the distribution function. However, we will concentrate
on a limited sample of the top $1000$ ranked ($k \le 1000$) journals in Set I and II.
To check if the overall functional form of the distribution
remains invariant with time, we rescaled the quantities by their averages,
and the curves appear to collapse into some universal function
irrespective of the year.
It may be noted that for small ranks, the citation is almost independent of rank implying
a cluster of journals with comparable citations that  occupy the top  positions
(Fig. \ref{fig:rank}A, B). Hence  the curves are  fitted for $k> 10$ by $f(x) \sim x^{-b_n}$ 
and we find that $b_n = 0.70(2)$. 
Similarly, for the IF, the scaled data is seen to fit to the same form with
an exponent $b_I= 0.54(1)$ (Fig. \ref{fig:rank}C, D). 
These exponents are called  the Zipf
exponents as they are obtained from the rank plots.
For SOCSCI, we find that  approximate power law fits are possible for the rank
plots,
the Zipf exponents  being 
$b_n= 0.70(2)$ 
and $b_I = 0.40(1)$  (Fig.~\ref{fig:rank}F, H).
\subsection{Correlation}
The  linear correlation coefficient is a measure of the strength of linear relation 
between two quantitative variables, say $x_i$ and $y_i$.
We use $R$ to represent the sample correlation coefficient:
\begin{equation}
\label{eq:correlation}
R=\frac{\sum_{i=1}^K (x_i-\bar{x})(y_i-\bar{y})}{\sqrt{{\sum_{i=1}^K (x_i-\bar{x})^2}\sum_{i=1}^K(y_i-\bar{y})^2}}
\end{equation}
Where $K$ is the number of individuals in the sample. We analyzed the data
using  $\log x_i$ and $\log y_i$ instead of $x_i$ and $y_i$ to include the general
case of $y$ having power-law dependence on $x$.
\begin{figure}[t]
\begin{center}
\includegraphics[width=8.5cm]{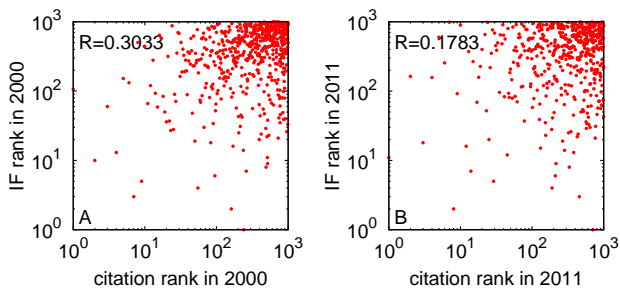}
\end{center}
      \caption{
      Rank according to  impact factor  versus rank according to citation for the journals common to Sets I and II,
shown for the years 2000 and 2011. The data is from SCI sets.}
  \label{fig:ifcites}
\end{figure} 
\begin{figure}
\begin{center}
\includegraphics[width=8.5cm]{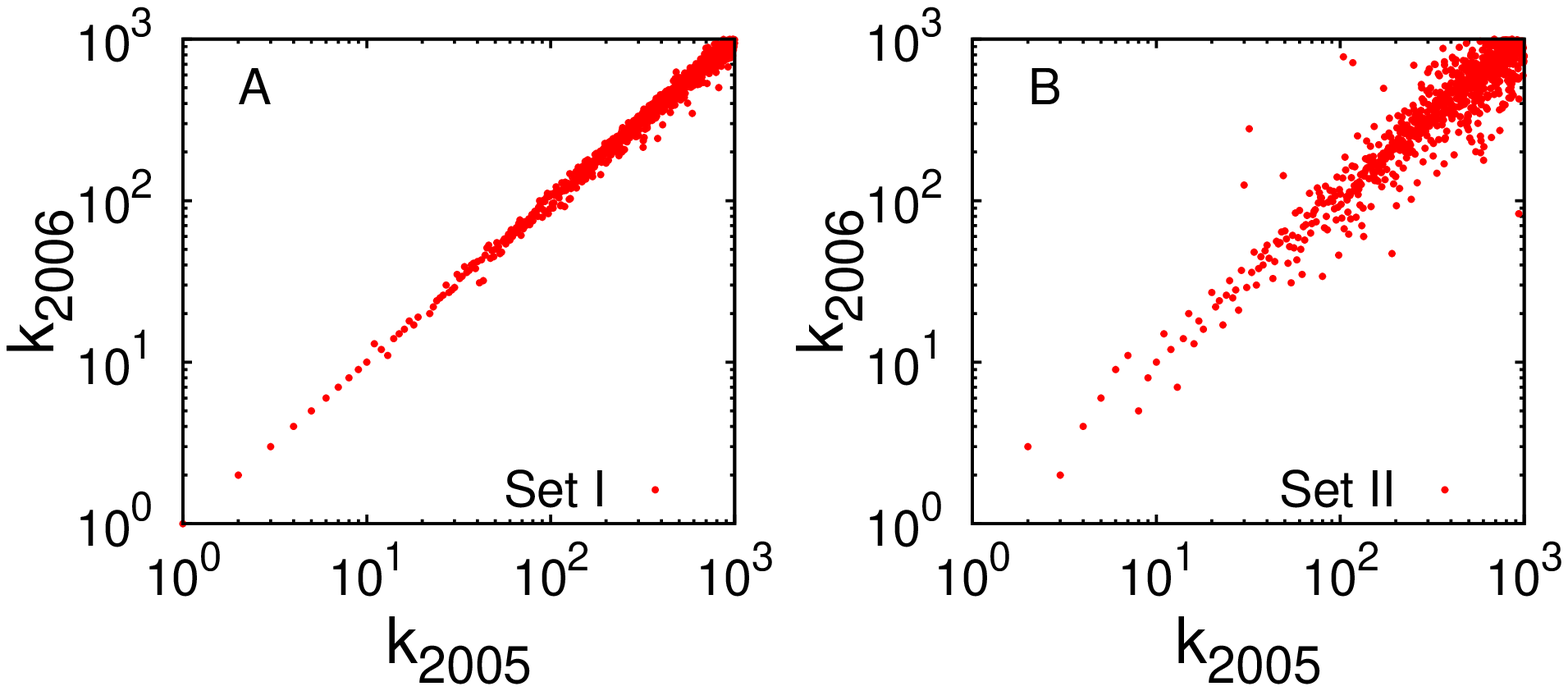}
    \includegraphics[width=8.5cm]{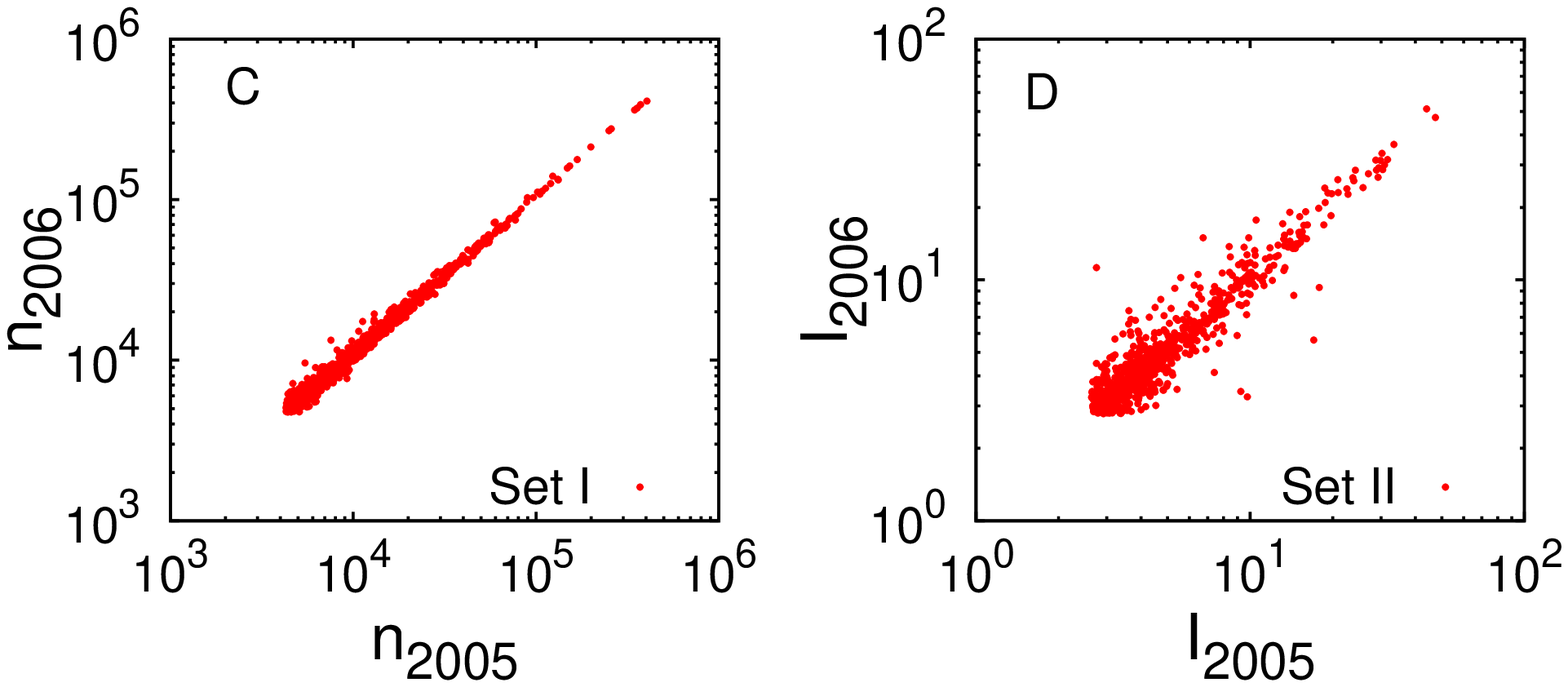}
\end{center}
      \caption{
      (A) Correlation between citation rank in the common set ($905$) of journals for 2005 and 2006, ranked by citations (Set I); $R \approx 0.996$.
    (B) Correlation between IF rank in the common set ($825$) of journals for 2005 and 2006, ranked according to annual citations (Set II); $R \approx 0.944$.
  (C) Correlation between annual citations $n$ in the common set  ($905$) of journals for 2005 and 2006 (Set I) ($R \approx 0.986$), and 
(D) Correlation between IF $I$ in the common set ($825$) of journals for 2005 and 2006 (Set II); $R \approx 0.923$.
 The data is from SCI sets.}
\label{fig:cor2}
 \end{figure}
\paragraph{Overlap and rank correlation of the sets: } 
We try to quantify how close the two data sets Set I and Set II are.
It is intuitively obvious that highly cited journals have higher IF and there
should be a reasonable overlap.
Fig.~\ref{fig:nvsi}, which  shows the AC vs IF `phase space'
- how Set I (ranked according to $n$)  and Set II (ranked according to $I$) sample different sets of a huge database -  
shows  also  the  overlap of the two sets. 
It is  interesting to find out how the  ranks according to AC and IF
are correlated. 
To calculate the   correlation  between AC and IF ranks of a journal,
we  plot the ranks of the common journals for two different years
in Fig.~\ref{fig:ifcites}. It is apparent that the ranks are quite 
uncorrelated (supported by low $R$ values). Both the overlap and rank correlation studies show that 
studying both Sets I and II are important,  as, first of all there are a lot of 
journals occurring  only in one set, and secondly, even when they occur in both sets, their 
positions (i.e.  ranks) in the two datasets are quite different.

\paragraph{Dynamic rank  correlations:}
Next we identify the journals and look at the scatter plots of AC (and 
IF) ranks in  two consecutive  years
to find out dynamic correlations, if any.  
We obtain the list of journals which occur  in Set I of both the years for the AC ranks 
and in Set II for IF ranks. The number of such journals is  larger when one considers AC 
ranks indicating that the position within the first 1000 journals is more stable when AC is considered. This 
is not surprising as citation data involves citation to all previously published  papers while IF is concerned with 
only recent papers.  As an example, if one considers the years 2005 and 2006, the number of common journals 
which occur in Set I is $905$ while for Set II it is $825$.
There is clear indication that for AC, the rank correlations are
much stronger (Fig.~\ref{fig:cor2}(A))  compared to IF rank correlations (Fig.~\ref{fig:cor2}(B)). 
Linear regression gives the values of correlation coefficients $R$ expectedly higher in case of citations.

\paragraph{Dynamic  correlation of actual values:}
Correlation coefficients for the 
actual values of IF and AC of the journals common in the sets for two different years were also calculated and 
they show identical trends. In fact the correlation coefficients  are very close when calculated   in terms of actual values 
 (Fig.~\ref{fig:cor2}(C) and (D)) 
and  ranks.
This indicates that small changes in the values will induce  small changes in ranks. 
The regression analysis for all possible pairs of years was done and discussed later in this paper.

\begin{figure}
\begin{center}
  \includegraphics[width=8.5cm]{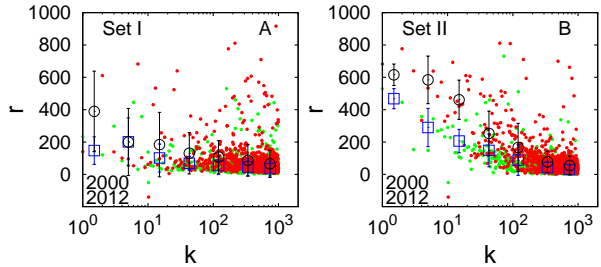}
\end{center}
    \caption{
    Rank ($k$) plots of the citation rates $r$ for the years 2000 and 2012 of the top $1000$ journals, ranked according to (A) citations (Set I),
    and (B) according to impact factors (Set II). 
    The average $r$ over rank bins are also shown ($\Box$ for 2000 and $\circ$ for 2012) along with their error bars.
The data are from SCI sets.}
\label{fig:crr}
 \end{figure}

\paragraph{Correlation between different measures:}
The annual citation rate (CR) $r$ has been plotted against both AC and IF ranks using the data in Set I and Set II.
It is interesting to note that 
{citation rates are less sensitive to citation ranks}  
compared to IF ranks, for which it 
shows a sharper decreasing trend (Fig.~\ref{fig:crr}). 
We have plotted data from different years to distinguish fluctuations across years from trends across sets.
\begin{figure}
\begin{center}
      \includegraphics[width=8.5cm]{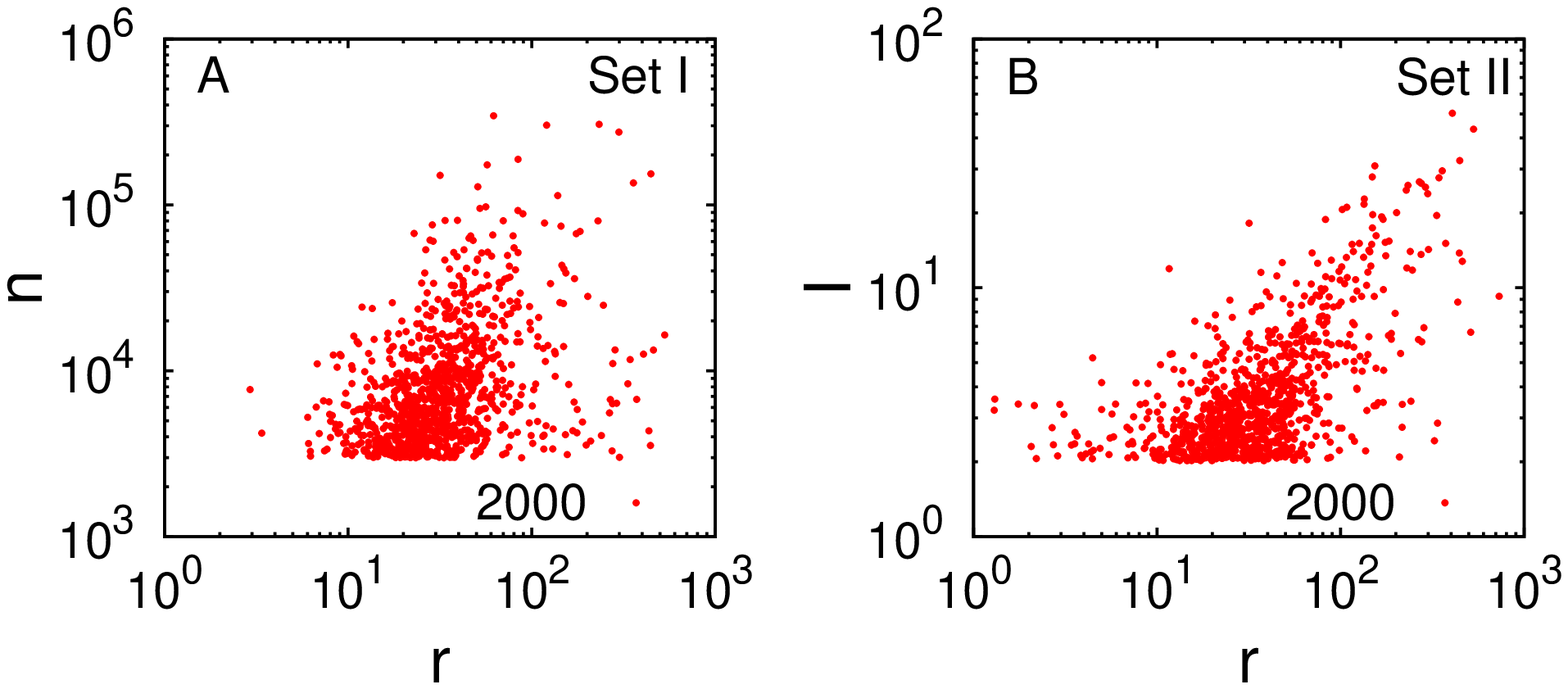}
    \includegraphics[width=8.5cm]{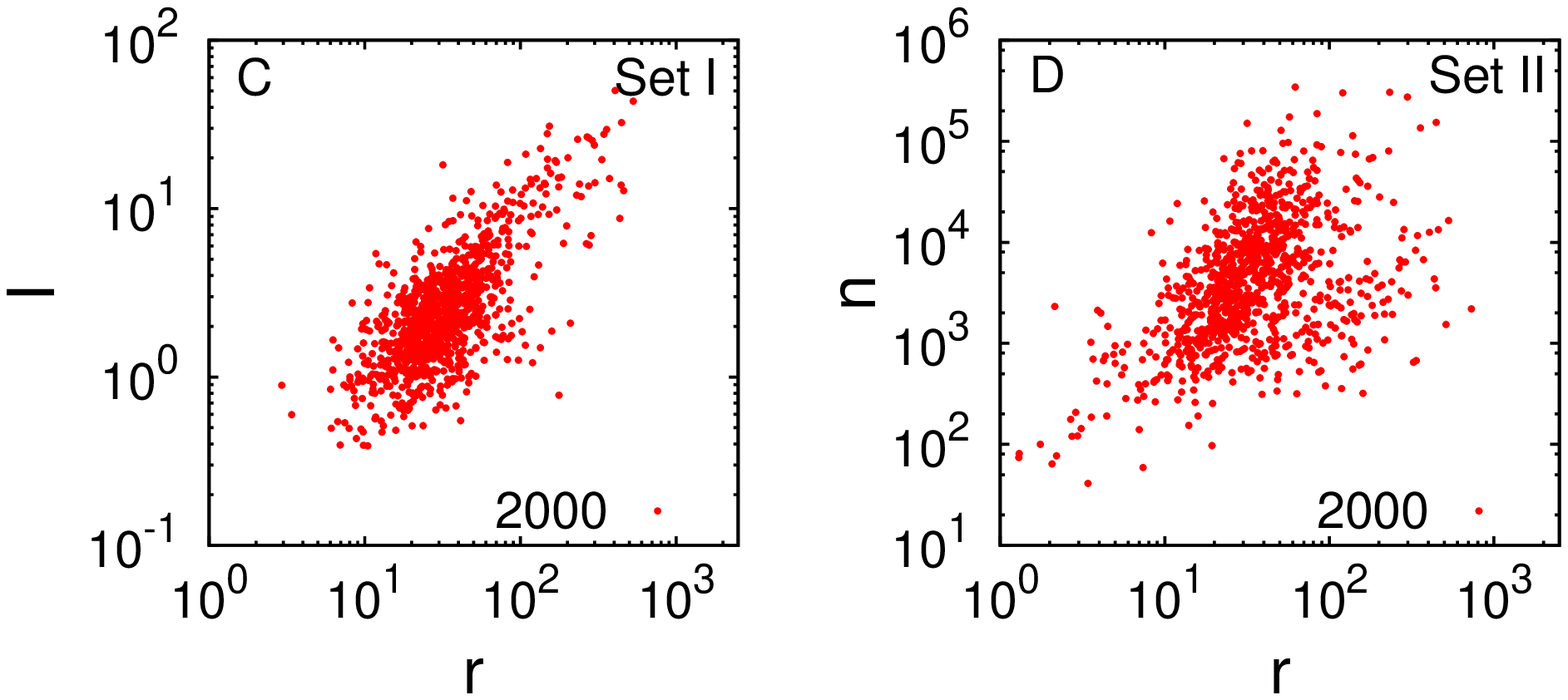}
\end{center}
\caption{
(A) Citation $n$ vs citation rate $r$ plot  for the top $1000$ journals, ranked according to citations (Set I). 
   (B) Impact factor $I$ vs citation rate $r$ plot for the top $1000$ journals, ranked according to impact factor (Set II).
Both plots are for 2000 for SCI data.
(C) IF $I$ vs citation rate $r$ plots for the top $1000$ journals, ranked according to citations (Set I). 
(D) Citation $n$ vs citation rate plots for the top $1000$ journals, ranked according to impact factor (Set II). 
Calculated value of correlation coefficients are (C) $0.7187$ and (D) $0.4526$.
}
\label{fig:crr1}
 \end{figure}

Correlation of values of $r$ with $n$ and $I$  are 
shown in   
Fig.~\ref{fig:crr1}, 
The correlations can be calculated from both set I and set II.
The plot of $r$ with   $n$ and $I$  made using the data in 
 set I and set II respectively 
are shown in Fig.~\ref{fig:crr1}(A) and (B).   
 A lower cutoff exists in these plots (as also for the data shown in 
 Fig.~\ref{fig:crr}) and therefore one should not calculate the 
correlation coefficients for these data.
Correlations coefficients can be calculated when one 
plots $r$ against $n$ taken from set II or $r$ against $I$ taken from set I (Fig.~\ref{fig:crr1}(C) and (D)).
 Both plots clearly show that CR and AC are much less correlated as was indicated from the rank plot (Fig.~\ref{fig:crr}) and the values of  the correlation
coefficients obtained from the `indirect' plots confirm this.

That AC and CR do not show considerable  correlation
indicates that the fluctuations in the number of publications  in different journal
is considerable.
 IF and $r$ are both scaled by the number of publications and  therefore they show more correlation.

\subsection{Distribution of  annual citations, IF and annual citation rate: nature of their tails}
\label{sec:tails}
\begin{figure}
\begin{center}
\includegraphics[width=8.5cm]{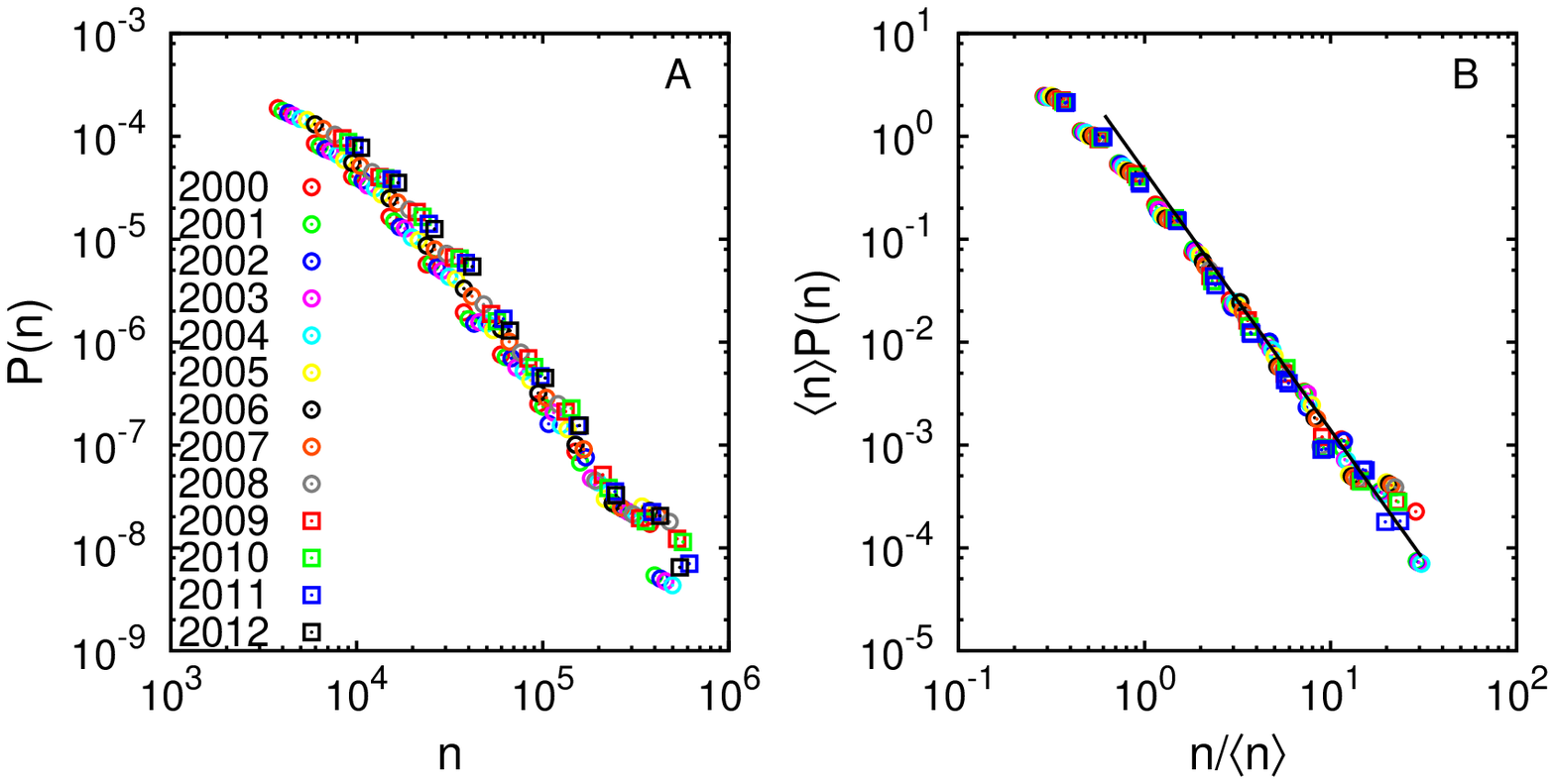}
\includegraphics[width=8.5cm]{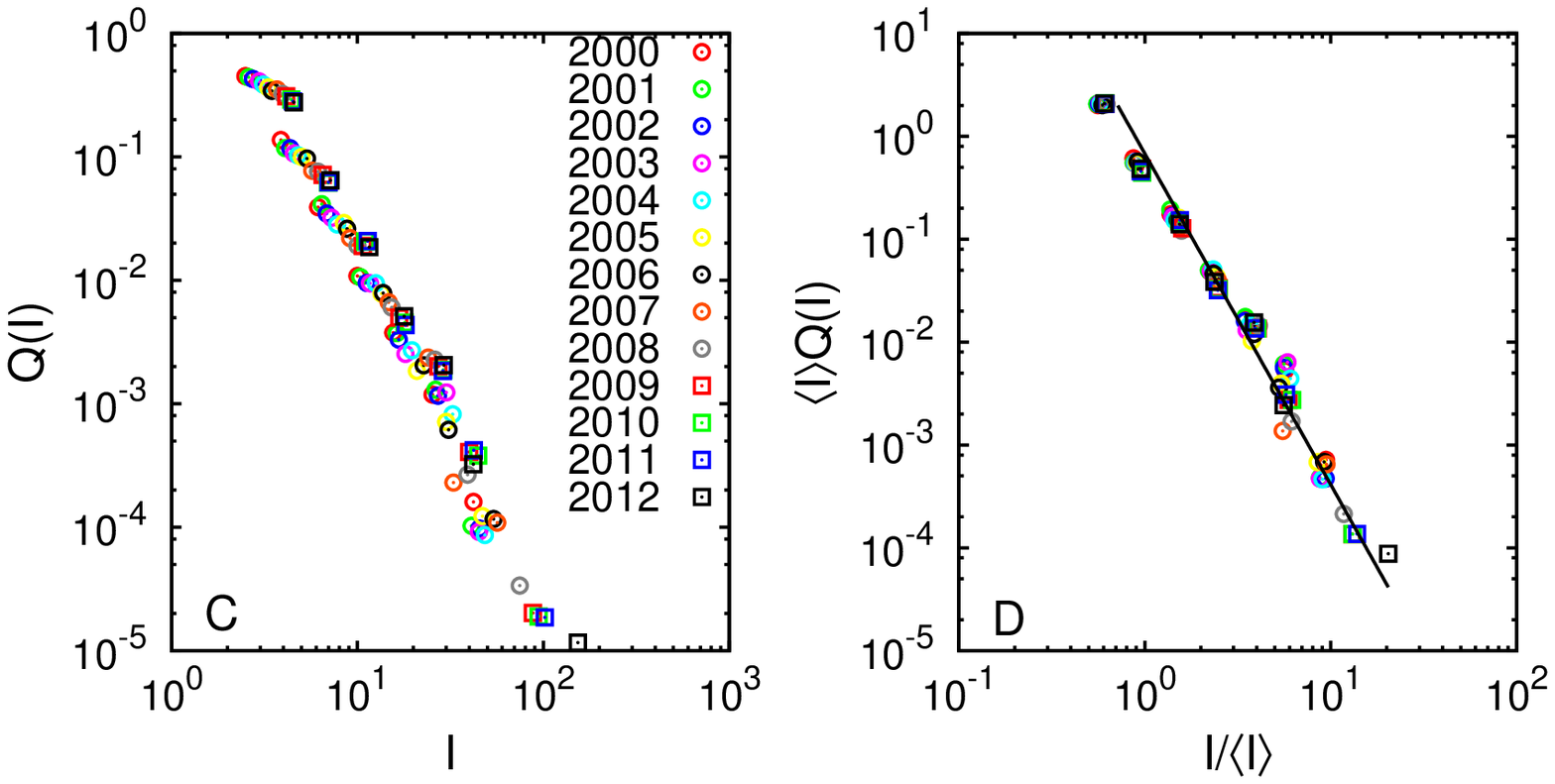}
\includegraphics[width=8.5cm]{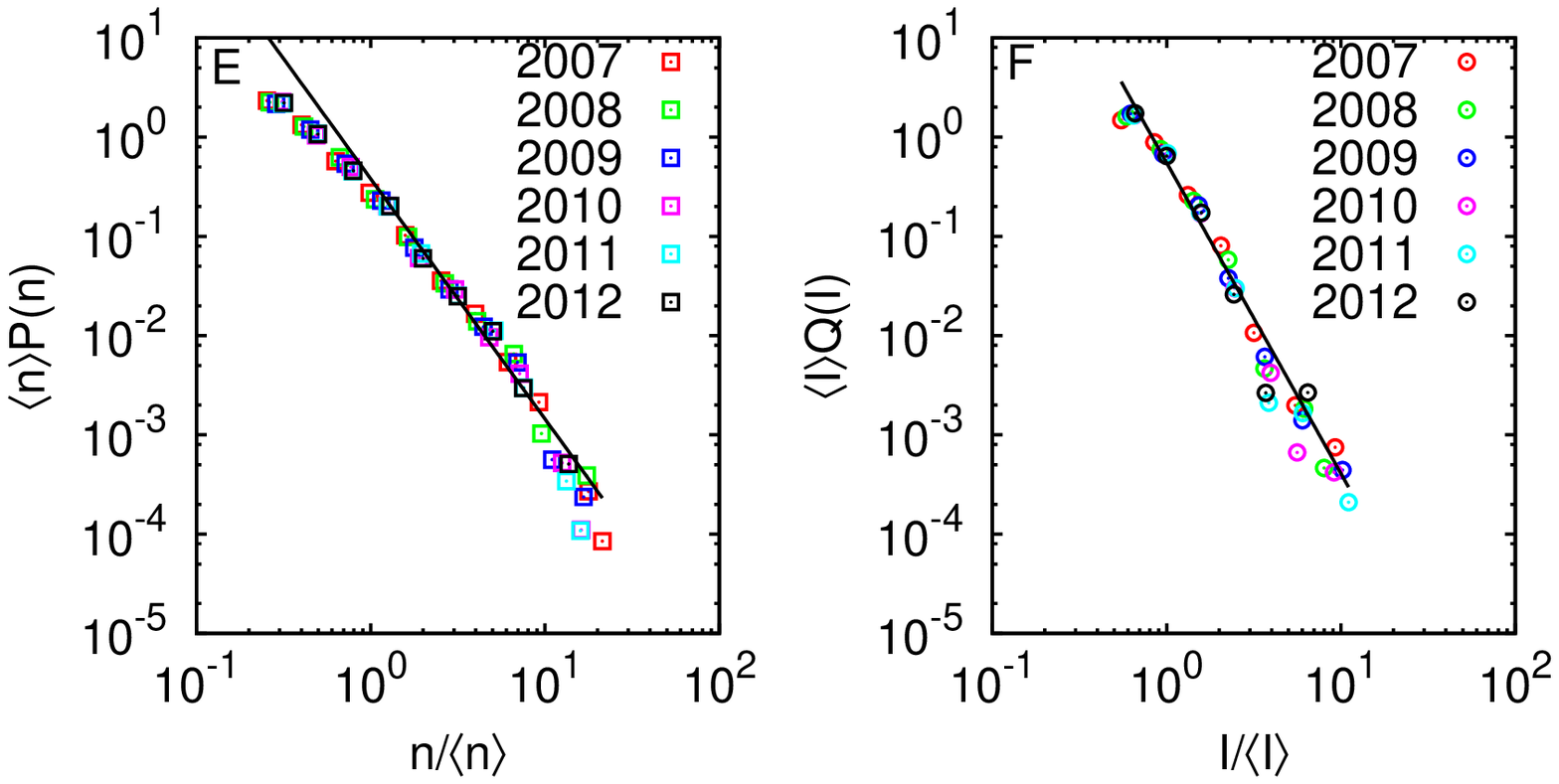}
\end{center}
     \caption{
     (A) Probability distribution of annual citations $P(n)$
   (B) scaling collapse of the same,  for the top $1000$ journals 
ranked according to citations (Set I).
It fits fairly well to a power law $B x^{-\gamma_n}$, and 
the straight line has slope $\gamma_n = 2.52$ in the log-log plot; 
  (C) probability distribution of impact factor $Q(I)$ 
 (D) scaling collapse of the same,
 for the top $1000$ journals 
ranked according to impact factor (Set II). 
The straight line has slope $\gamma_I = 3.16$.
The data are from SOCSCI sets:
(E) Scaling collapse of probability distribution of annual citations $P(n)$
 for the top $1000$ journals ranked according to citations (Set I);
 power law fit with $\gamma_n = 2.42$;
 (F) Scaling collapse of probability distribution of annual citations $P(n)$
 for the top $1000$ journals ranked according to impact factor (Set II);
  power law fit with $\gamma_I = 3.50$.     The data are from SCI sets.
}
\label{fig:dist}
 \end{figure}

First we  investigate  the nature of the tail of the distribution of
annual citations $P(n)$ (Fig.~\ref{fig:dist}(A)) and  impact factors $Q(I)$ (Fig.~\ref{fig:dist}(C)) from Set I and II.
The citation and impact distributions do not represent the entire dataset as 
only the highest ranked journals are considered. Therefore only the tail of the distribution 
can be obtained here. 
The distribution of annual citations and  impact factors showed monotonic 
decays, the tail of which can be fitted to power law forms.
There is a  lower cutoff in the annual citation and impact factor
and no local peak is expected.
The plots showed excellent scaling collapse over years when 
in general for any  probability distribution $X(x)$, $X(x) \langle x\rangle$ is
plotted against $x / \langle x \rangle$.
For annual citations, the distribution at lower values hint towards a lognormal
but we concentrate on the behavior of the `tail' of the distribution, which is a power law~\cite{Golosovsky:2012}.
The power law exponents   (also called the Pareto exponents) $\gamma_n$ and $\gamma_I$ 
are $2.52(1)$ and $3.16(1)$ respectively.
The Zipf exponent $b$  obtained from a rank plot  is related to the Pareto exponent 
$\gamma$  obtained  from the probability  distribution by $\gamma = 1+1/b$~\cite{Clauset:2009}.
Using the values of $b$ reported in \ref{rank}, the values of $\gamma$   
are  $2.42$ and $2.85$ respectively for AC and IF, which compare quite well to
the values obtained directly from the plots of the distributions
(Fig. \ref{fig:dist}).
The same is true for the SOCSCI data (expected Pareto exponents are $2.42$ and 
$3.50$ from the Zipf plots while best fit values are $2.32(2)$ and $3.13(2)$), where, 
however,  it is apparent that there  
may be some corrections to power law scaling.

\begin{figure} 
\begin{center}
\includegraphics[width=8.5cm]{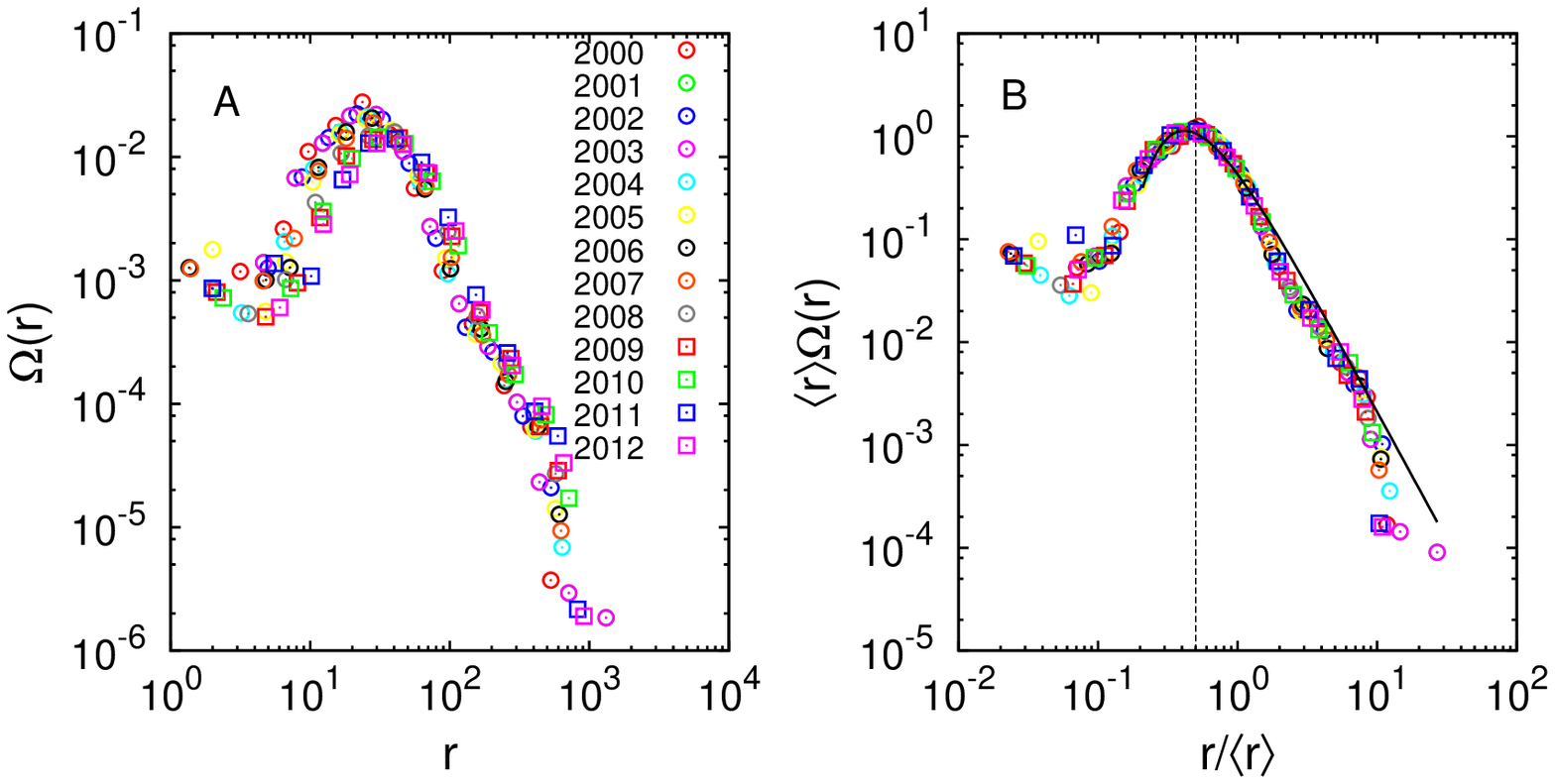}
    \includegraphics[width=8.5cm]{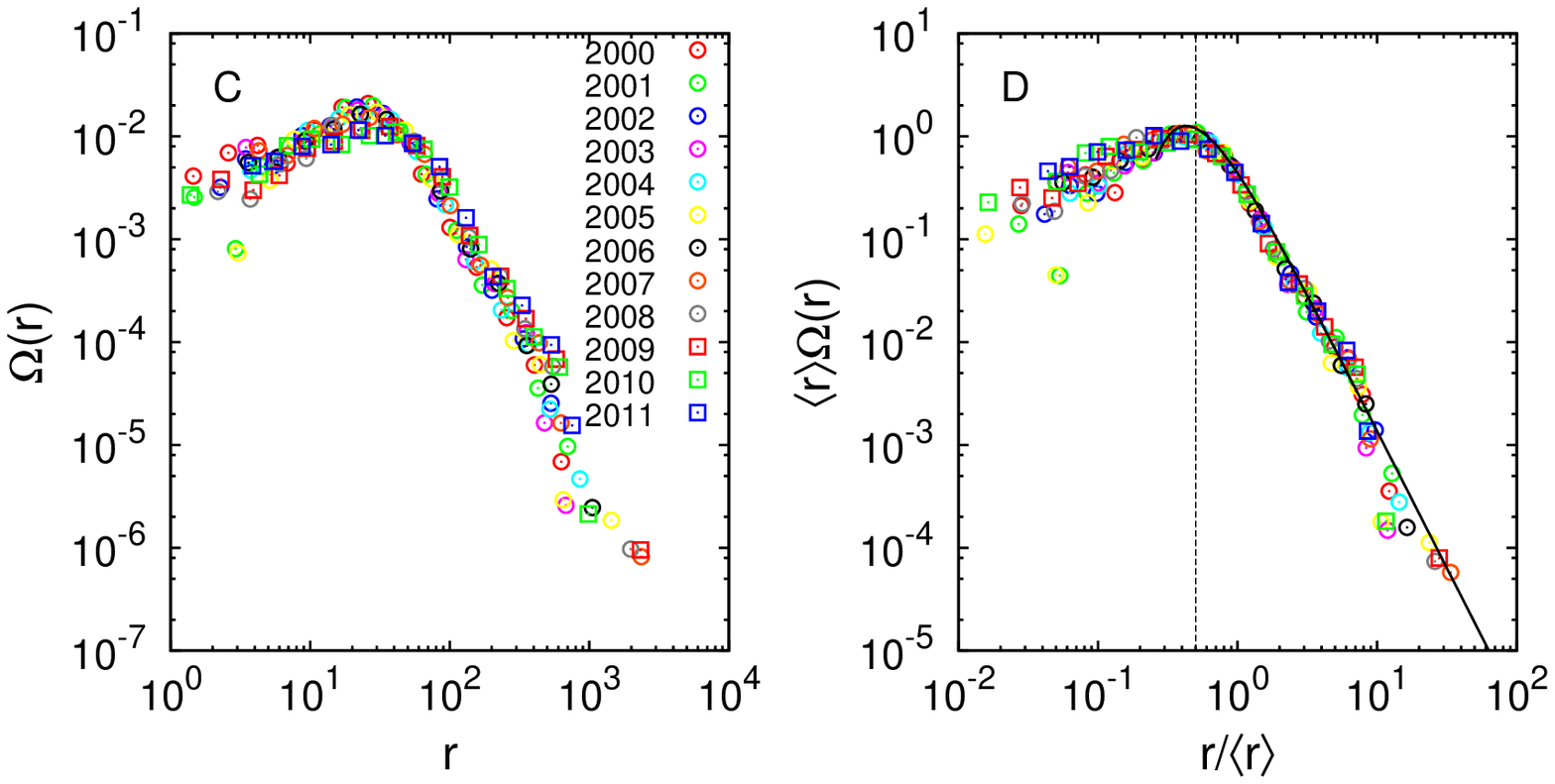}
    \includegraphics[width=8.5cm]{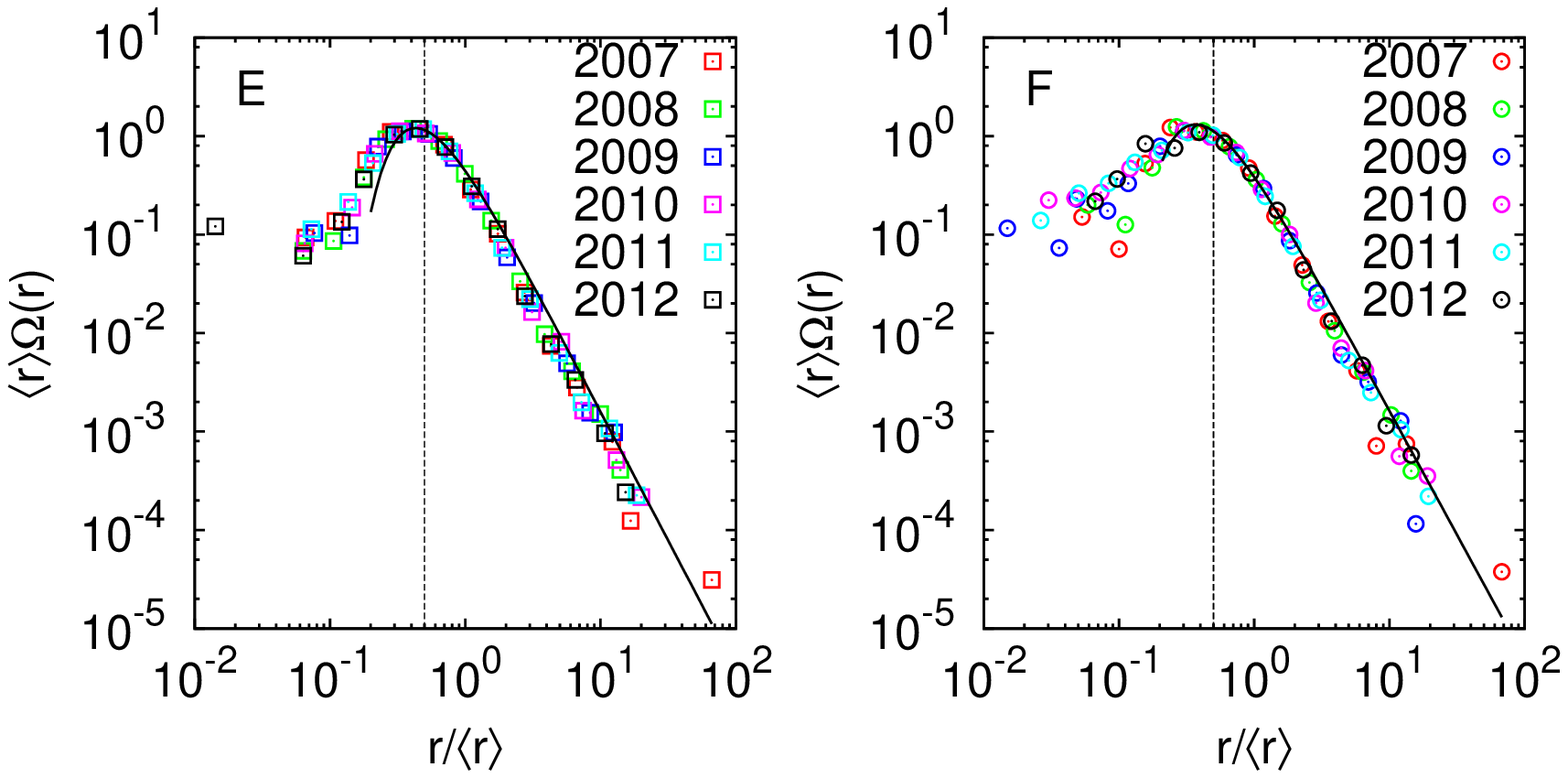}
\end{center}
     \caption{
     (A) Probability distribution of annual citation rate $\Omega(r)$ from Set I (citation ranked),
    (B) scaling collapse of the same. It fits well to a log-Gumbel distribution $\frac{1}{b}e^{-(z+e^{-z})}$, where $ {z}=\frac {x-a}{b}$ giving $a=-0.5385$ and $b=0.6677$.
 (C) probability distribution of annual citation rate $\Omega(r)$ from Set II (IF ranked), and
(D) scaling collapse of the same ($a=-0.5711$ and $b=0.5986$)
The data are from SCI sets for the top $1000$ journals.
The data are from SOCSCI sets for the top $1000$ journals:
(E) Scaling collapse of Set I with $a=-0.546$ and $b= 0.6969$;
(F) Scaling collapse of Set II with $a=-0.6691$ and $b=0.6554$.
The data from SCI sets for the top $1000$ journals.}
\label{fig:citrate}
 \end{figure}

The probability
distributions $\Omega(r)$  of the newly proposed quantity, the annual citation rate $r$ (Fig~\ref{fig:citrate}A-D) 
computed from Set I and Set II share
similar characteristics, although they differ by the absolute values of their fluctuations.
Distributions of annual citation rates $r$ are non-monotonic, compared to $I$ and $n$, 
they have a peak but eventually decay at large $r$ (approximately as $r^{-3}$).
The distribution is also consistently narrow with respect to that of $I$ and $n$.
The curves for successive years also showed 
excellent scaling collapse, when scaled with the averages.
The non-monotonic behavior for both sets is characterized by
a prominent peak at $r^* = r/\langle r \rangle \approx 0.5$.
The appearance of a most probable value indicates that most of
the journals are likely have a similar value of annual citation rate
which is approximately half the sum of the annual citation rates
of all journals.
The corresponding plots for SOCSCI are shown in Fig~\ref{fig:citrate}E-F,
and these sets are consistent with the above behavior.

The tail of the distribution $\Omega(r)$ seems to fit to a power law, but in order
to account for the maximum part of the data, including the non-monotonicity and the peak,
we proposed a fitting using a log-Gumbel function.
We performed the Kolmogorov-Smirnov test to evaluate
the quality of the fit: we have sorted the data points into increasing order  for the year 2000.
We compare the empirical pattern and theoretical estimated value.
The largest error is  $0.2325$ for sample size $N=12$ (SET I).  For level of significance $s=0.20$,
the critical value of daviation is $0.295$ for this  sample size. 
The largest error is $0.1917$ for sample size $N=14$ (SET II) and  the critical value is $0.274$.
For the SOCSCI set: The largest error is
$0.3153$ for sample size $N = 14$ (Set I). For level
of significance $s = 0.05$, the critical value of deviation is $0.349$ for this sample size. 
The largest error is $0.2732$ for sample size $N = 14$ (Set II)
and For level of significance $s = 0.20$ the critical value is $0.274$. Since our estimated
daviation is less than the critical value, we use the
log-Gumbel distribution for fittings.

We had seen earlier that CR is much less correlated with the AC
compared to the IF, but as far as distributions are concerned, one does not notice this 
difference except perhaps in minute details.

\section{Summary  and discussions}
We studied  various static and dynamic properties related to citations of journals, in
 science (SCI) and social science (SOCSCI) databases. 
The datasets that have been used were constructed on the basis of ranking according to 
annual citation and impact factor, but we find that a third measure, 
introduced as the annual citation rate (CR) gives interesting insights into the  data. 
For example, Set I and II provide only the data for the high ranked journals 
so that it is possible to study only the tail of the distribution of the quantity according to which
the corresponding set is ranked. 
However,  for the CR, one can obtain a distribution which also shows the existence of  a
peak independent of the set used. 
Additionally, we find an  universal property with respect to the position of the peak.
A few  important observations made in the present work are summarized  below:
\begin{enumerate}
 
 \item 
Scaling the number of citations (either to recent papers  or all papers published in a journal) by 
the total number of (annual) publications  is crucial as the  properties depend on this scaling.
This is indicated by the difference in the  behavior of   annual citations (AC), impact factors (IF)  and the  citations rate (CR)
 --  the CR values show considerably larger correlations with the IF ranks and values.

\item Distribution of AC and IF show general behavior similar to each other --
the highest values of both can be fitted to power law forms  (Pareto)  (Fig.~\ref{fig:dist}) 
with exponents fairly  consistent with those 
obtained from the rank (Zipf)  (Fig.~\ref{fig:rank}) plots for the Science journals.
For Social Science journals, this is also fairly consistent.

\item The probability distribution of the annual citation rate CR has an universal feature -- it shows
a maximum at half of the average rate, irrespective of how the journals are ranked.
This feature is completely unique from annual citations or impact factors whose
probability distributions decay monotonically. 
This  holds true for both Science and Social Science journals (Fig.~\ref{fig:citrate}).
Distributions of both sets can be fitted to log-Gumbel distribution forms. 
 
\end{enumerate}

A relevant question is 
whether  authors  should  choose higher IF journals for submission. 
This choice definitely depends on many  factors, the content and impact of the problem and the results  
reported in  the paper being the most important  ones.  
Usually researchers have some idea which are the  journals where the acceptance probability 
is higher for a particular paper. So a set of a few suitable journals   are considered for submission.
For jobs or promotion of researchers,   evaluation might be made on the basis of   
the number of citations to their papers ($h$ index is a 
popular measure)  and/or by the journal IF where 
their papers have been published. The second measure is actually quite useless as 
only current IFs are quoted while their papers might have been published more than two years before.
In fact, in more recent times, the policy  is \textit {not} to consider the IF for jobs or promotion.

IF is, however,  definitely important from the journal's viewpoint for advertising/commercial purpose 
and journals do need to develop policies to improve it. 
Authors still  prefer journals  with larger IF when submitting papers (irrespective of job or other prospects) 
among  the suitable journals
merely because a comparative measure is available although  such a measure might be misleading as already
emphasised. A higher IF provides a psychological edge and the tendency to submit manuscripts 
to a suitable journal with higher IF   is like a  preferential attachment.  However, this does not
necessarily lead to a rich get richer effect for the journal. This is because  if a submitted paper is accepted, it does not
guarantee that it will lead to a larger IF in the  coming two years. Rather, if it fails to get cited within the next
two years (there is a finite possibility of this, in fact many papers do not get cited at all within 
a finite time~\cite{Rousseau:1994,Egghe:2000}),  the IF is bound to suffer.
Hence a strategy to increase IF may be to reduce the number of accepted papers and already 
such efforts have been noted   in the data.
We found that the average IF as a function of number of articles  
published  in different journals shows a weak decreasing trend (for        
large number of articles) in 2012
 as compared to 2000 to support this idea, however, the  data contains   
large fluctuation and it is better not to make any  definitive statement based on this data.
Another method to generate larger IF is to increase readership
by advertising in many forms. 
A larger readership ensures greater exposure of papers and probability of citations; the reverse is also true to some extent.

\bibliographystyle{unsrt}
\bibliography{cita.bib}

\end{document}